# Diffusion time dependence of microstructural parameters in fixed spinal cord

Sune Nørhøj Jespersen[1,2,*], Jonas Lynge Olesen[1,2], Brian Hansen[2], Noam Shemesh[3]

Author affiliations

[1]Center of Functionally Integrative Neuroscience (CFIN) and MINDLab, Department of Clinical Medicine, Aarhus University, Aarhus, Denmark.

[2]Department of Physics and Astronomy, Aarhus University, Aarhus, Denmark.

[3]Champalimaud Neuroscience Programme, , Lisbon, Portugal

[*]Corresponding author:

Sune Nørhøj Jespersen

CFIN/MindLab and Dept. of Physics and Astronomy, Aarhus University

Nørrebrogade 44, bygn 10G, 5. sal

8000 Århus C

Denmark

Cell: +45 60896642

E-mail: sune@cfin.au.dk

## Abstract

Biophysical modelling of diffusion MRI is necessary to provide specific microstructural tissue properties. However, estimating model parameters from data with limited diffusion gradient strength, such as clinical scanners, has proven unreliable due to a shallow optimization landscape. On the other hand, estimation of diffusion kurtosis (DKI) parameters is more robust, and its parameters may be connected to DKI parameters, given an appropriate biophysical model. However, it was previously shown that this procedure still does not provide sufficient information to uniquely determine all model parameters. In particular, a parameter degeneracy related to the relative magnitude of intra-axonal and extra-axonal diffusivities remains. Here we develop a model of diffusion in white matter including axonal dispersion and demonstrate stable estimation of all model parameters from DKI in fixed pig spinal cord. By employing the recently developed fast axisymmetric DKI, we use stimulated echo acquisition mode to collect data over an unprecedented diffusion time range with very narrow diffusion gradient pulses, enabling finely resolved measurements of diffusion time dependence of both net diffusion and kurtosis metrics, as well as model intra- and extra-axonal diffusivities, and axonal dispersion. Our results demonstrate substantial time dependence of all parameters except volume fractions, and the additional time dimension provides support for intra-axonal diffusivity to be larger than extra-axonal diffusivity in spinal cord white matter, although not unambiguously. We compare our findings for the time-dependent compartmental diffusivities to predictions from effective medium theory with reasonable agreement.

## Introduction

Diffusion weighting is the main method for imparting microstructural sensitivity in MRI. In this imaging mode, tissue water diffusion can be assessed over typical measurement times of a few tens of milliseconds. This timescale corresponds to diffusion lengths on the order of a few micrometers largely coinciding with characteristic length scales in tissue (cell sizes, etc.). Diffusion MRI (dMRI) can therefore be used to probe tissue microstructure indirectly, effectively allowing interrogation of structures orders of magnitude below nominal imaging resolution. Membranes, organelles and other components of tissue microstructure act as obstacles to diffusion and cause it to deviate from Gaussian diffusion. Therefore, quantifying nongaussian diffusion will in principle lead to improved microstructural sensitivity in diffusion MRI. Even so, the dMRI signal decay in most biological tissues remains rather unremarkable (Yablonskiy and Sukstanskii, 2010) and no decay curve feature yields quantitative microstructure parameters directly. Therefore, the combination of dMRI with biophysical modelling is key to characterizing tissue microstructure. Unfortunately, estimating parameters of biophysical models has proved rather challenging, due to i) the complex nature of biological tissue often requiring sophisticated models with many free parameters, and ii) the aforementioned lack of features in the diffusion signal decay, which allows it to be fit well by many different models. More frustratingly, a fundamental degeneracy has been demonstrated so that the signal can be fit equally well by the same model with different parameter values (Jelescu et al., 2015; Jelescu et al., 2016). Hence, estimating all parameters of complex microstructural models robustly is likely to require comprehensive data sets including very high diffusion weighting, which is feasible only on specialized scanners and mostly ex vivo (Alexander et al., 2010; Assaf and Basser, 2005; Assaf et al., 2008; Jespersen et al., 2007a; Jespersen et al., 2010).

Many currently used models in diffusion MRI share a similar picture of the underlying neural tissue, namely that of neurites as long hollow cylinders (e.g. (Assaf and Basser, 2005; Assaf et al., 2008; Fieremans et al., 2011; Jelescu et al., 2015; Jensen et al., 2016; Jespersen et al., 2007a; Jespersen et al., 2010; Kroenke et al., 2004; Novikov et al., 2016b; Veraart et al., 2016b; Zhang et al., 2012)). This picture was therefore recently referred to as the "standard model" of diffusion in neural tissue (Novikov et al., 2016a).  While estimation of model parameters in this class of models is feasible with very high b-value diffusion data sets, strategies directed towards robust parameter estimation from data of clinical grade have surfaced only recently. One such approach, pioneered in (Fieremans et al., 2011), aims to take advantage of the stability of fitting the diffusion signal to the diffusion kurtosis imaging signal expression.

Diffusion kurtosis imaging (DKI) (Jensen et al., 2005) is a popular method allowing to capture the dMRI signal's leading deviations from Gaussian diffusion. DKI builds on the so-called cumulant expansion (Kampen, 2007; Kiselev, 2011; Risken, 1984) of the log signal yielding signal terms weighted by increasing powers of the diffusion weighting strength b. The cumulant expansion is a very general framework, and so DKI does not rely on strong assumptions about the tissue from which the dMRI signal is obtained. Consequently, DKI applies very broadly, almost regardless of the true tissue microstructure. In fact, the applicability of DKI is limited in practice only by the requirement of not too high b-values ($bD \leq 3$), which makes the technique well suited for standard clinical diffusion sequences/hardware. In order to reveal microstructural detail, however, DKI must be combined with biophysical modelling relating the terms of the cumulant expansion to specific tissue properties (Fieremans et al., 2011; Hui et al., 2015; Jespersen et al., 2012; Novikov et al., 2016b). The caveat of this approach is that in general, the diffusion and kurtosis tensors do not supply sufficient information to unambiguously determine all parameters of the standard model (Novikov et al., 2016b). In regions of highly aligned fibers, however, this degeneracy is isolated to a choice between 2 solutions (branches), corresponding to intra-axonal diffusivity being either larger than or smaller than extra-axonal diffusivity (Fieremans et al., 2011). These 2 solutions, i.e. sets of microstructural parameters, both give the exact same diffusion and kurtosis tensors, and therefore fit the observed diffusion signal decay equally well in the applicable range of DKI. Since this range roughly corresponds to the range typically probed on clinical scanners, this degeneracy is fundamentally related to the difficulty of estimating parameters of microstructural models from clinical data in general (Jelescu et al., 2016). In order to resolve this redundancy in the choice of microstructural parameters, additional information is necessary. This can be provided by higher diffusion weighting (Novikov et al., 2016b) as has been observed in the past (Jespersen et al., 2010; Jespersen et al., 2007b), or by acquiring "orthogonal information" as supplied for example by extended diffusion encoding sequences such as double diffusion encoding (Cory et al., 1990; Jespersen et al., 2013; Koch and Finsterbusch, 2009; Lawrenz et al., 2010; Mitra, 1995; Shemesh and Cohen, 2011; Shemesh et al., 2016; Shemesh et al., 2010a; Shemesh et al., 2010b), or by isotropic diffusion encoding and "fat b-tensors" (Dhital et al., 2015; Lampinen et al., 2017; Lasic et al., 2014; Szczepankiewicz et al., 2015; Topgaard, 2017). Alternatively, varying the diffusion time may also provide sufficient information to identify the correct branch choice.

Time-dependent diffusion is another hallmark of nongaussian diffusion in biological tissue (Latour et al., 1994; Mitra et al., 1993; Novikov et al., 2014; Novikov et al., 2016a), and is being increasingly studied as a window into tissue microstructure. Several observations of time-dependent diffusivity have been reported (Aggarwal et al., 2012; Baron and Beaulieu, 2014; Does et al., 2003; Fieremans et al., 2016; Horsfield et al., 1994; Kershaw et al., 2013; Kunz et al., 2013; Portnoy et al., 2013; Tanner, 1979; Van et al., 2014; Wu et al., 2014), but the literature is more sparse concerning time dependence of kurtosis parameters (Pyatigorskaya et al., 2014). This is presumably due to the higher demands for SNR and data compared to DTI, reflecting the increased number of parameters (22 versus 7).

Here we use a recent advance, axisymmetric DKI combined with fast DKI (Hansen et al., 2016b), to acquire a comprehensive dataset comprising diffusion weighted images over a broad range of diffusion times, allowing a detailed analysis of both diffusivity and kurtosis time dependence. The data is subsequently combined with a biophysical model of diffusion MRI based on the standard model to yield time dependence of intra-and extra-axonal diffusivity of both solutions. We show that this indeed provides more evidence for one branch over the other branch, corresponding to extra-axonal diffusivity higher than intra-axonal diffusivity, based on the expected physical behavior. Our results for the time dependence of compartmental diffusivities are further compared to predictions (Burcaw et al., 2015; Novikov et al., 2012) based on effective medium theory.

## Theory

### Biophysical modelling

We employ a variant of the 'standard model' mentioned above designed in particular for spinal cord white matter. Thereby, our framework will be familiar to most readers and our analysis can be considered quite general and connected to an existing body of literature (Fieremans et al., 2011; Jelescu et al., 2015; Jespersen et al., 2010; Jespersen et al., 2007b; Kroenke et al., 2004; Novikov et al., 2016b; Veraart et al., 2016b; Zhang et al., 2012). Our model is comprised of two non-exchanging, Gaussian compartments representing the axonal and extra-axonal spaces. Axons are approximated by zero radius sticks characterized by an orientation distribution function, and diffusion in both individual axons and the surrounding extra-axonal space is approximated by a Gaussian propagator. Hence, the model is characterized by an intra-axonal diffusivity $D_a$, extra-axonal radial and axial diffusivities, $D_{e,\perp}$ and $D_{e,\parallel}$, and the axonal water fraction $f$, in addition to the fiber orientation distribution function (fODF) $P(\hat{u})$. The diffusion signal $S$ is then generally given by (Novikov et al., 2016b; Reisert et al., 2017):

$$S(b,\hat{n}) = \int d\hat{u} P(\hat{u}) \Big( f \exp(-bD_a(\hat{u} \cdot \hat{n})^2) + (1-f)\exp(-bD_{e,\perp} - b(D_{e,\parallel} - D_{e,\perp})(\hat{u} \cdot \hat{n})^2) \Big) \qquad (1)$$

where we have suppressed the diffusion time dependence, as it will analyzed at a later stage (see below).

We emphasize that the assumption of Gaussian diffusion within the compartments is a main limitation of the standard model. This is especially critical when the analysis relies on estimating net kurtosis from either single diffusion encoding (Fieremans et al., 2011; Hansen et al., 2017; Novikov et al., 2016b) or isotropic diffusion encoding (Dhital et al., 2015; Lampinen et al., 2017), because the net kurtosis has contributions from both variance in diffusivities as well as intracompartmental kurtosis, which may be of comparable size – this is independent of diffusion weighting b. However, it is plausible that for long diffusion times, spins average our local microstructural fluctuations to render an effective medium with approximately Gaussian diffusion (Novikov and Kiselev, 2010). Likewise, the standard model also ignores other complexities of neural tissue such as microcirculation, edema, myelin debris etc., which may contribute a non-negligible signal fraction under some circumstances. In such cases, poor fitting or unrealistic parameter values may result, and clearly more validation is necessary to define the standard model's range of applicability.

As treated in detail in (Jespersen et al., 2007b), the fiber ODF and the kernel in Eq. (1) decouple when expressed in terms of the spherical harmonics expansion:

$$S(b,\hat{n}) = \sum_{l=0,2,4\ldots} \sum_{m=-l}^{l} s_{lm}(b) Y_{lm}(\hat{n}) = \sum_{l=0,2,4\ldots} \frac{4\pi}{2l+1} \sum_{m=-l}^{l} k_l(b) p_{lm} Y_{lm}(\hat{n}) \qquad (2)$$

where the $p_{lm}$ are the spherical harmonics expansion coefficients of the fiber ODF $P(\hat{u})$, and $s_{lm}(b) = k_l(b) p_{lm}$ the spherical harmonics expansion coefficients of the signal with

$$k_l(b) = f C_l(bD_a) + (1-f) e^{-bD_{e,\perp}} C_l\left(b(D_{e,\parallel} - D_{e,\perp})\right), \qquad (3)$$

where the functions $C_l$ are defined similarly to (Jespersen et al., 2007b),

$$C_l(x) = \frac{2l+1}{2} \int_{-1}^{1} dz P_l(z) e^{-xz^2}, \qquad (4)$$

and can be given explicitly for a given $l$, for example (Jespersen et al., 2007b):

$$C_0(\mathbf{x}) = \sqrt{\frac{\pi}{4x}} \mathrm{erf}\left(\sqrt{x}\right)$$

$$C_2(\mathbf{x}) = -\frac{5\left(\sqrt{\pi}(2x-3)\mathrm{erf}\left(\sqrt{x}\right) + 6e^{-x}\sqrt{x}\right)}{8x^{3/2}}$$

$$C_4(\mathbf{x}) = \frac{9\left(3\sqrt{\pi}(4x(x-5)+35)\mathrm{erf}\left(\sqrt{x}\right) - 10e^{-x}\sqrt{x}(2x+21)\right)}{64x^{5/2}}$$

$$C_6(\mathbf{x}) = \frac{-65\sqrt{\pi}\left(8x^3 - 84x^2 + 378x - 693\right)\mathrm{erf}\left(\sqrt{x}\right) - 546e^{-x}\sqrt{x}(4(x+5)x+165)}{256x^{7/2}}$$

(5)

With the current measurements focusing on spinal cord tissue, we make the further approximation that the fibre ODF is axially symmetric around some axis $\hat{c}$ (Zhang et al., 2012). In this case, Eq. (2) simplifies to the Legendre expansion via the selection of the $m = 0$ components only

$$S(b,\hat{n}) = \sum_{l=0,2,4\ldots} P_l(\hat{c} \cdot \hat{n}) k_l(b) p_l \tag{6}$$

Here, $P_l$ are the Legendre polynomials, and $p_l \equiv \sqrt{4\pi/(2l+1)}\, p_{l0}$ (in the system where $\hat{c} = \hat{z}$) are expansion coefficients of the axially symmetric fODF

$$P(\hat{u}) = \sum_{l=0,2,4\ldots} \frac{2l+1}{4\pi} P_l(\hat{c} \cdot \hat{u}) p_l. \tag{7}$$

Despite their difference to the rotational invariants introduced in (Novikov et al., 2016b; Reisert et al., 2017) in the general case, their absolute values are proportional by a factor of $4\pi$ in the axially symmetric case. The most obvious approach, terminating the expansion at some $l = L$ in Eq. (6), and estimating the parameters $f$, $D_a$, $D_{e,\perp}$, $D_{e,\parallel}$, $p_2$, $p_4, \ldots p_L$ through nonlinear fitting of the signal to Eq. (6) proved to be unstable for our data. The same was the case for the procedure based on estimating the signal expansion coefficients $s_{lm}(b)$ for each shell separately by linear least squares, followed by estimation of the fODF and kernel expansion coefficients by nonlinear fitting to $s_l(b) = k_l(b) p_l$ using Eq. (3).

We therefore pursue a different approach taking advantage of the relatively stable fitting of diffusion kurtosis imaging, and the existence of fast kurtosis acquisition and estimation methods (Hansen et al., 2013; Hansen et al., 2014; Hansen et al., 2016a; Hansen et al., 2016b) based on axisymmetric kurtosis imaging (see below). Within the general model in Eq. (1), analytical expressions for diffusion and kurtosis tensors in terms of model parameters are readily found (Jespersen et al., 2012), as well as all higher order moments (Novikov et al., 2016b). In the case of an axially symmetric fiber ODF, the diffusivity D and kurtosis tensor W only depend on the projection $\xi = \hat{c} \cdot \hat{n}$ of the diffusion direction onto the symmetry axis $\hat{c}$, and we find the explicit expressions

$$D(\xi) = fD_a h_2(\xi) + (1 - f)\left( (D_{e,\parallel} - D_{e,\perp}) h_2(\xi) + D_{e,\perp} \right) \tag{8}$$

$$W(\xi)\overline{D}^2 = 3\left( fD_a^2 h_4(\xi) + (1-f)\left( D_{e,\perp}^2 + (D_{e,\parallel} - D_{e,\perp})^2 h_4(\xi) + 2D_{e,\perp}(D_{e,\parallel} - D_{e,\perp})^2 h_2(\xi) \right) - D(\xi)^2 \right) \tag{9}$$

where the functions $h_2$ and $h_4$ are related to moments of the fibre ODF:

$$h_2(\xi) = \int d\hat{u} P\,(\hat{u})(\hat{u} \cdot \hat{n})^2 = \frac{1}{3} + \frac{2}{3} p_2 P_2(\xi)$$
$$h_4(\xi) = \int d\hat{u} P\,(\hat{u})(\hat{u} \cdot \hat{n})^4 = \frac{1}{5} + \frac{4}{7} p_2 P_2(\xi) + \frac{8}{35} p_4 P_4(\xi) \tag{10}$$

Due to the assumed axial symmetry, Eqs. (8) – (9) supply 5 independent equations corresponding to 2 rotational invariants for D and 3 for W (Hansen et al., 2016c). Hence it is not possible to determine all 6 unknown model parameters $f$, $D_a$, $D_{e,\perp}$, $D_{e,\parallel}$, $p_2$, $p_4$ without further information or simplification (Novikov et al., 2016b).

Approximating axonal fibers to be completely parallel as in the original WMTI (Fieremans et al., 2010), ignoring dispersion altogether, reduces the number of degrees of freedom by 2 by setting $p_2 = p_4 = 1$. The explicit analytical expressions of directional kurtosis metrics in this case become (Hansen et al., 2017):

$$D_\perp = (1 - f)D_{e,\perp}$$
$$D_\parallel = fD_a + (1 - f)D_{e,\parallel}$$
$$W_\perp \overline{D}^2 = 3f(1 - f)D_{e,\perp}^2$$
$$W_\parallel \overline{D}^2 = 3f(1 - f)(D_a - D_{e,\parallel})^2 \tag{11}$$

with explicit solutions:

$$f = \left( 1 + 3D_\perp^2 / W_\perp \overline{D}^2 \right)^{-1}$$
$$D_{e,\perp} = D_\perp / (1 - f)$$
$$D_{e,\parallel} = D_\parallel - \frac{2}{3}\frac{f}{1-f}\left( D_\perp + \eta \sqrt{\frac{15(1-f)}{4f} \overline{D}^2 \overline{W} - 5D_\perp^2} \right)$$
$$D_a = D_\parallel - \frac{2}{3}\left( D_\perp - \eta \sqrt{\frac{15(1-f)}{4f} \overline{D}^2 \overline{W} - 5D_\perp^2} \right) \tag{12}$$

Subsequent refinement of WMTI was shown to accommodate some fiber dispersion, however it was not modeled explicitly and relied on approximate estimates for $f$ and $D_a$ (Fieremans et al., 2011).

We propose to model dispersion explicitly by employing a one-parameter family of axially symmetric ODFs. In this case, the 2 parameters $p_2$ and $p_4$ contain only one (common) degree of freedom, and the system of equations (8) and (9) can therefore be solved exactly for the 5 independent microstructural parameters. A commonly used choice is the Watson distribution $p(\hat{u}) \propto \exp(\kappa(\hat{u} \cdot \hat{c})^2)$ with concentration parameter $\kappa$, yielding

$$
\begin{aligned}
p_2 &= \frac{1}{4}\left( \frac{3}{\sqrt{\kappa}\,\mathrm{F}(\sqrt{\kappa})} - 2 - \frac{3}{\kappa} \right) \\
p_4 &= \frac{1}{32\kappa^2}\left( 105 + 12\kappa(5 + \kappa) + \frac{5\sqrt{\kappa}(2\kappa - 21)}{\mathrm{F}(\sqrt{\kappa})} \right)
\end{aligned}
\tag{13}
$$

where F is Dawson's function (Abramowitz et al., 1972). Other axially symmetric parameterizations are of course also possible (Reisert et al., 2017). We then solve for the model parameters $f$, $D_a$, $D_{e,\perp}$, $D_{e,\parallel}$, $\kappa$ by matching Legendre expansion coefficients of Eq. (8) to the measured ones, essentially equivalent to LEMONADE (Novikov et al., 2016b), yielding the system of equations:

$$
\begin{aligned}
3D_0 &= fD_a + (1-f)(2D_{e,\perp} + D_{e,\parallel}) \\
\frac{3}{2}D_2 &= p_2\left[ fD_a + (1-f)(D_{e,\parallel} - D_{e,\perp}) \right] \\
D_2^2 + 5D_0^2(1 + \frac{W_0}{3}) &= fD_a^2 + (1-f)\left[ 5D_{e,\perp}^2 + (D_{e,\parallel} - D_{e,\perp})^2 + \frac{10}{3}D_{e,\perp}(D_{e,\parallel} - D_{e,\perp}) \right] \\
\frac{1}{2}D_2(D_2 + 7D_0) + \frac{7}{12}W_2D_0^2 &= p_2\left[ fD_a^2 + (1-f)\left( (D_{e,\parallel} - D_{e,\perp})^2 + \frac{7}{3}D_{e,\perp}(D_{e,\parallel} - D_{e,\perp}) \right) \right] \\
\frac{9D_2^2}{4} + \frac{35}{24}W_4D_0^2 &= p_4\left[ fD_a^2 + (1-f)(D_{e,\parallel} - D_{e,\perp})^2 \right]
\end{aligned}
\tag{14}
$$

where $D_0$ and $D_2$ are the Legendre expansion coefficients of D in Eq. (8), and $W_0$, $W_2$ and $W_4$ are the Legendre expansion coefficients of W in Eq. (9). In terms of the measured diffusion and kurtosis parameters we have:

$$D_0 = \overline{D} = \frac{1}{3}(2D_\perp + D_\parallel)$$

$$D_2 = \frac{2}{3}(D_\parallel - D_\perp)$$

$$W_0 = \overline{W} \tag{15}$$

$$W_2 = \frac{1}{7}(3W_\parallel + 5\overline{W} - 8W_\perp)$$

$$W_4 = \frac{4}{7}(W_\parallel - 3\overline{W} + 2W_\perp)$$

where subscripts $\perp$ and $||$ identify radial and axial diffusion and kurtosis components, corresponding to $\xi = 0$ and $\xi = 1$, respectively, in Eqs. (8) and (9). After using Eq. (13) to relate $p_2$ and $p_4$, Eqs. (14) can be solved analytically (Novikov et al., 2016b). Practically, we solve the first four equations by expressing the volume fraction and all diffusivities in terms of $p_2$, and then determine $\kappa$ by solving the last equation numerically. Due to the appearance of squared diffusivities, two solutions (branches) exist (Fieremans et al., 2011; Jelescu et al., 2016; Novikov et al., 2016b). We report the behavior of both, labelled by branch choice, $\eta = \pm 1$, corresponding to $\eta = 1$ for $4 - \sqrt{40/3} < (D_a - D_{e,\parallel})/D_{e,\perp} < 4 + \sqrt{40/3}$, and $\eta = -1$ otherwise (Novikov et al., 2016b). With the current data, these 2 branches are found to fulfil $D_a > D_{e,\parallel}$ for $\eta = 1$ and $D_a < D_{e,\parallel}$ for $\eta = -1$.

All the procedures described above estimate microstructural parameters for each specific diffusion time. The time dependence of the intra-axonal and extra-axonal diffusivities is subsequently compared to the effective medium theory predictions (Burcaw et al., 2015; Fieremans et al., 2016; Novikov et al., 2014; Novikov and Kiselev, 2010):

$$D_a(t) \sim (D_a(\infty) + \frac{c_1}{\sqrt{t}})$$

$$D_{e,\parallel}(t) \sim (D_{e,\parallel}(\infty) + \frac{c_2}{\sqrt{t}}) \tag{16}$$

$$D_{e,\perp}(t) \sim D_{e,\perp}(\infty) + \frac{c_3}{t}\ln(t/t_c)$$

where $c_1$, $c_2$ and $c_3$ are constants. The correlation time $t_c$ corresponds to the time taken to diffuse across the correlation length. Effective medium theory predicts the functional form of the approach to the longtime asymptote to depend on the type of structural disorder probed by the spins. Conceptually, the range of structural correlations determines the rate at which diffusion approaches its longtime Gaussian fixed point by averaging out (coarse graining) local fluctuations, but does not directly address the microstructural underpinnings of the disorder. Both functional forms above are the result of Poissonian disorder, in one dimension ($1/\sqrt{t}$) and in two dimensions ($\ln(t)/t$). Note that effective medium theory predictions apply only asymptotically when $t \gg t_c$. Since the correlation time is unknown a priori, we fit from a cutoff time to the largest diffusion time and monitor the error as the cutoff time is decreased. The final cutoff time is fixed to be a point at which the normalized error increases steeply (see Figure 7 in Results). In contrast to the correlation time $t_c$, the cutoff time does not have any obvious physical interpretation by itself.

*Axisymmetric DKI*

Axisymmetric DKI is a recently invented technique allowing accurate estimation of all DKI parameters, under the assumption of axially symmetric fiber ODFs, on the basis of nine diffusion directions at two non-zero b-values, in addition to a *b*=0 image (Hansen et al., 2016c). It is based on replacing the general diffusion ( $D_{ij}$ ) and kurtosis tensors ( $W_{ijkl}$ ) in the DKI signal expression:

$$
\begin{aligned}
\log S(b,\hat{n}) / S_0 &= -b n_i n_j D_{ij} + \frac{1}{6} b^2 \overline{D}^2 n_i n_j n_k n_l W_{ijkl} = -bD(\hat{n}) + \frac{1}{6} b^2 \overline{D}^2 W(\hat{n}) \\
&= -bD(\hat{n}) + \frac{1}{6} b^2 D(\hat{n})^2 K(\hat{n})
\end{aligned}
\tag{17}
$$

with their axially symmetric versions

$$
\mathrm{D} = D_\perp \mathrm{I} + (D_\parallel - D_\perp)\hat{c}\hat{c}^T
\tag{18}
$$

$$
\mathrm{W} = \frac{1}{2}(10W_\perp + 5W_\parallel - 15\overline{W})\mathrm{P} + W_\perp \mathrm{I} + \frac{3}{2}(5\overline{W} - W_\parallel - 4W_\perp)\mathrm{Q}.
\tag{19}
$$

Here, $\hat{n}$ is the diffusion gradient direction, $\hat{c}$ is the (unknown) axis of symmetry and subscripts $\perp$ and || identify radial and axial diffusion kurtosis components. Furthermore, $\mathrm{I}$ is the rank 2 identity matrix, and the tensors

$$
\begin{aligned}
\mathrm{P}_{ijkl} &= c_i c_j c_k c_l \\
\mathrm{Q}_{ijkl} &= \frac{1}{6}(c_i c_j \delta_{kl} + c_i c_k \delta_{jl} + c_i c_l \delta_{jk} + c_j c_k \delta_{il} + c_j c_l \delta_{ik} + c_k c_l \delta_{ij}) \\
I_{ijkl} &= \frac{1}{3}(\delta_{ij}\delta_{kl} + \delta_{ik}\delta_{jl} + \delta_{il}\delta_{jk})
\end{aligned}
\tag{20}
$$

were introduced. It was previously shown that axisymmetric DKI affords reliable estimates of radial, axial and mean diffusion and kurtosis, even in regions where actual axial symmetry is unlikely (Hansen et al., 2016c).

Due to the reduced number of parameters of axisymmetric DKI, 8 compared to 22 for general DKI, the 199 protocol (Hansen et al., 2013; Hansen et al., 2016a; Hansen et al., 2016b) provides sufficient data for the estimation of all parameters in Eqs. (8) and (11) using nonlinear least squares as demonstrated for the axisymmetric case in (Hansen et al., 2017).

## Methods

### Imaging

Fixed porcine spinal cord was imaged with a stimulated echo diffusion imaging sequence (EPI) on a 16.4T Bruker Aeon Ascend magnet interfaced with an AVANCE IIIHD console, and equipped with a micro5 probe with gradients capable of producing up to 3000 mT/m in all directions. Spinal cord specimens were obtained from domesticated adult pigs, and the tissues were immersed in 4% PFA roughly 30 minutes post-mortem. The samples were kept in PFA for at least 48h prior to washing with PBS, and placement in a 10 mm NMR tube filled with Fluorinert (Sigma Aldrich, Portugal). The sample was then imaged using a 10 mm birdcage coil tuned for 1H. A measure of the SNR was computed as the mean signal in the spinal cord divided by the mean signal in a noise region. For nominal b=0, SNR varied between 46 and 76 over all diffusion times, and for nominal b=6 ms/$\mu$m$^2$, it varied between 27 and 46.

Six b=0 images were acquired, as well as six b-value shells from 0.5 to 3 ms/$\mu$m$^2$. Each shell was encoded using the nine directions for the 199 DKI scheme (Hansen et al., 2016a). Such data sets were acquired for 57 diffusion times ($\Delta$) ranging from $\Delta$ = 6 ms to 350 ms with 4 averages each. Across these data sets, the following imaging parameters were kept constant: TE=16 ms, TR=7.5 s, in-plane resolution 140 $\mu$m x140 $\mu$m, slice thickness = 2.2 mm, and gradient pulse width $\delta$ = 1.15ms. Sample temperature was maintained at 25°C throughout using air flow.

The raw data are freely available for download at http://cfin.au.dk/cfinmindlab-labs-research-groups/neurophysics/data/.

### Postprocessing

The raw images were denoised (Veraart et al., 2016a) and corrected for Gibbs ringing (Kellner et al., 2016) before the subsequent analysis. Axial, radial, and mean diffusion and kurtosis values were estimated with axisymmetric DKI (Hansen et al., 2016b) using Levenberg-Marquardt based nonlinear least squares (software available at http://cfin.au.dk/cfinmindlab-labs-research-

groups/neurophysics/software/). All analysis was performed in Matlab (The Mathworks, Natick, MA, USA) using actual b matrices including cross terms with imaging gradients (Lundell et al., 2014). Throughout, our analysis focused on seven regions of interest (ROIs) corresponding to known spinal cord white matter (WM) tracts as identified in (Ong and Wehrli, 2010). The ROIs are shown on an FA image in Fig. 1. The outlined tracts are: A=dorsal corticospinal tract, B=Fasiculus Gracilis, C=Fasiculus Cuneatis, D=Reticulospinal tract, E=Rubrospinal tract, F=Spinothalamic tract, G=Vestibulospinal tract. In these regions of relatively well aligned axons, we extract biophysical parameters characterizing the WM bundles from the kurtosis and diffusion tensors estimated from the data. Fitting biophysical model parameters as function of diffusion time to effective medium theory was also done in Matlab using nonlinear fitting with the Levenberg-Marquardt algorithm.

## Results

*DKI Metrics*

Figure 2 shows time dependence of diffusion tensor (top row) and kurtosis metrics (bottom row) over the entire acquired range of diffusion times 6-350 ms for all seven ROIs (A-G). All parameters show substantial time variation, and most of them appear not to have plateaued fully even at the largest time point (350 ms), possibly with the exception of $D_\perp$, $K_\perp$ and $\overline{W}$ for a few of the ROIs. The diffusivities all decrease monotonically. This is expected on physical grounds, with the rate of decay decreasing as the diffusion time increases. Radial diffusivity decreases relatively more than axial diffusivity, more than a factor of two for all of the ROIs. Axial diffusivity is substantially larger than radial diffusivity at all times, and there are clear differences among the ROIs. The magnitudes of the axial diffusivities vary across the ROIs and this order is preserved over all diffusion times except for ROIs C and E. This behavior is also seen for the radial diffusivities and the mean diffusivities although here the ROI curves are ordered differently. This indicates that the tortuosity may vary greatly among these tracts. The kurtosis metrics are more noisy than the diffusion metrics, especially radial kurtosis, where some of the ROIs (especially D,G) show considerable scatter for diffusion times exceeding ~90 ms. Axial kurtosis and mean kurtosis show an intriguing nonmonotonic behavior for some of the ROIs, with an initial increase in kurtosis followed by a decrease. All ROIs have decreasing axial and mean kurtosis for sufficiently long times, whereas radial kurtosis initially increases, and possibly plateaus for longer times (A,C,E,F). Note that mean kurtosis is independent from radial and axial kurtosis, in contrast to the case of the diffusivities, where mean diffusivity can be computed from axial and radial diffusivity.

*Model parameters*

We first demonstrate that the estimated model parameters provide an accurate and precise approximation to the measured signal. In Figure 3, the acquired signal in each voxel is plotted versus the computed signal, i.e. one point per voxel from each of the diffusion weighted images (b-values, directions, and diffusion times), with one graph for each ROI. Perfect agreement would thus result in all of the points lying on the line $y = x$. The observed scatter around this line is small and approximately evenly distributed above and below, indicating a good fit to the data. Since the estimated model parameters precisely reproduce the diffusion and kurtosis tensors, this figure is equivalent to a graph of the DKI signal versus the measured signal, and shows that the DKI approximation is adequate.

In Figure 4 we show the behavior of the compartmental diffusivities as function of diffusion time for all ROIs and for both of the two solution branches. The top row shows intra-axonal diffusivity, middle row shows extra axonal axial diffusivity, and bottom row shows extra-axonal radial diffusivity. We describe them in turn.

*Intra-axonal diffusivity* $D_a$: The plus branch estimates (left) decrease smoothly as function of diffusion time, whereas the minus branch (right) shows as substantially more noisy behavior, and even a tendency for increasing values as function of diffusion time for some of the ROIs. The intra-axonal diffusivities of the plus branch are relatively high, approaching or even exceeding for one data point the bulk diffusivity of water $\approx 2.3$ μm²/ms (Holz et al., 2000) at 25° Celsius for the lowest diffusion times in some of the ROIs. Eventually, the diffusivities decrease roughly by 30-40% compared to their initial values. The relative magnitude of the intra-axonal diffusivities across the ROIs is mostly the same as for the net axial diffusivities. The analysis was repeated for the case of fully aligned fibers (Eq. (12)), and here the tendency for the diffusivities to increase as a function of diffusion time for the minus branch was even more pronounced.

*Extra-axonal axial diffusivity* $D_{e,\parallel}$: For the plus branch (left), $D_{e,\parallel}$ is smaller than the corresponding intra-axonal diffusivities by a factor of typically 3 or more depending on time and ROI, but the ROI order is similar. In the minus branch (right), estimates are upwards of a factor eight larger than the corresponding intra-axonal diffusivities, but in all cases remain below the free water diffusivity. For the plus branch, the relative decrease from the initial value ranges from 13% to 76%, and in contrast to the intra-axonal case, the extra-axonal axial diffusivities appear to have more or less plateaued for diffusion times larger than about 200 ms. This is in contrast to the minus branch, where the diffusivities seem to decay over the entire range of diffusion times (20% to 30%). The plus branch is perhaps slightly noisier than the minus branch here.

*Radial extra-axonal diffusivity* $D_{e,\perp}$: the radial extra-axonal diffusivities behave more similarly for the 2 branches, especially for small diffusion times. At the larger diffusion times, the minus branch generally decays slightly below that of the plus branch. For diffusion times above 100 ms the plus branch again seems noisier than the minus branch. At very early times, the plus branch starts out from values close to those of the plus branch axial diffusivities, consistent with the isotropic bulk diffusion value being approached. This is in contrast to the minus branch, where the radial diffusivities are less than approximately half of the axial diffusivities. As the longest diffusion times are approached, the minus branch radial extra-axonal diffusivities reach values that are 40%-50% of their initial values, whereas the relative decay for the plus branch is in the range of 34% to 43%. The order of the ROIs is mostly the same as for the net radial diffusivities.

The axonal water fraction, *f*, is shown in Figure 5. This parameter is seen to be largely independent of diffusion time in both branches, as would be expected. Typical values are 0.5-0.6 for the plus branch, and 0.2-0.4 for the minus branch. For diffusion times below approximately 75 ms, there appears to be a small systematic increase/decrease in its estimated value as function of diffusion time for the plus/minus branch, presumably due to shortcomings of the assumptions underlying the analysis.

The estimated axonal orientation dispersion is plotted in terms of the concentration parameter $\kappa$ of the Watson distribution in Figure 6. For more intuitive interpretation, the corresponding angular dispersion is reported on the right-hand y-axis, computed as $\arccos\left(\sqrt{\langle(\hat{c}\cdot\hat{u})^2\rangle}\right) = \arccos\left(\sqrt{1/3 + 2/3\,p_2}\right)$, where the average $\langle\cdot\rangle$ is over the fODF $P(\hat{u})$. For the plus branch (left), the dispersion decreases as function of diffusion time, starting at around 29° and ending between 15° and 19°, where the dispersion plateaus somewhat. For the minus branch on the other hand, dispersion hits the imposed maximum of 50 for the great majority of the points. Raising the upper bound to 100 results in almost all estimates becoming 100, with minimal effect on the remaining minus branch parameters.

*Comparison to EMT*

Based on the observations above, particularly the behavior in $D_a$ (Fig. 3) and dispersion (Fig. 7), we regard the plus branch as the most likely solution for our data. Therefore we base the following comparison to effective medium theory (EMT) on this branch. In Figure 7 we show an example of the EMT fitting procedure. In Fig. 8A, 3 fits for 3 different diffusion time ranges to $D_a^+(t)$ in the spinothalamic tract (F) are shown as solid curves of green, yellow and red. EMT predictions apply for sufficiently large diffusion times, and the selection of the appropriate range is illustrated in Figs. 8B and C. Fig. 8B plots mean square errors $E_n$ as function of cutoff time (lower limit of the diffusion time fitting range). The numbers have been normalized by the minimum mean square errors $E_{n,\min}$ achieved over all cutoff times. For very large cutoff times, the sum of squared errors is small due to good agreement to the asymptotic predictions, and few data points. As the cutoff time decreases, the number of data points increase and after a certain point the agreement to EMT deteriorates rapidly. The idea behind the selection of the cutoff time is to identify it with the point at which the error reaches an approximate plateau. In Fig. 8C, the rate of change $(\Delta E_n / E_{n,\min}) / \Delta t$ in $E_n / E_{n,\min}$, i.e. the slope of the data in the middle panel, is monitored as function of cutoff time, and is seen to rapidly increase in magnitude when 100 ms is approached from above. The intersection with the green, yellow and red lines identify the cutoff times corresponding to relative changes of 0.01, 0.05, 0.1.

Figure 8 shows the resulting fits to EMT for all ROIs (rows) and intra-axonal diffusivity (left column), axial extra-axonal diffusivity (middle column), and radial extra-axonal diffusivity (right column). The x-axes have been transformed according to the predicted functional behavior to render the theoretical behavior linear. The fits show good correspondence with the data for intra-axonal diffusivity in all ROIs. The agreement for the extra-axonal diffusivities is more variable across the ROIs, with good agreement in some and poorer agreement in others. However, the poor agreement in some of the ROIs seems to be more due to scatter than to a systematic deviation from the predictions.

The fitting parameters corresponding to the largest cutoff times (green lines) from all ROIs is summarized in Table 1 along with the histological measurements of axon volume fraction normalized to the volume of non-myelin tissue compartments and diameter in the identified WM tracts in rat spinal cord from (Nunes et al., 2017). Note that the axonal volume fraction values are more in line with the plus branch than the minus branch. Correlation lengths are on the cellular scale of a few micrometers, similar to the axon diameters, especially for extra axonal radial correlation length.

## Discussion

### Summary

In this study we presented multi-shell diffusion weighted data of the fixated pig spinal cord, with densely sampled diffusion times (57 in total) from 6 ms to 350 ms using very narrow diffusion pulses.

The data was analyzed in 7 white matter ROIs comprising 7 previously studied white matter fiber tracts. Using axisymmetric DKI, we were able to estimate radial and axial diffusivity, as well as radial, axial and mean kurtosis. All parameters demonstrated notable diffusion time dependence over most of the probed range. By matching these moments of the cumulant expansion to those computed from a biophysical model of diffusion in the spinal cord, we remapped the measurements to microstructural parameters, i.e. axonal volume fraction, axonal dispersion, intra-axonal diffusivity, and extra-axonal radial and axial diffusivity. This approach results in two a priori equally acceptable solutions (Fieremans et al., 2011; Novikov et al., 2016b), and the additional data dimension (diffusion time) allowed us to examine the behavior of both in detail, assuming that the branch choice is stable as a function of diffusion time. This is supported by the large difference in $D_a$ and $D_{e,\parallel}$ values for each branch at all diffusion times, since a change in the branch choice would lead to $D_a = D_{e,\parallel}$ at some diffusion time. Based on the observed time dependence of diffusivities and dispersion, we conclude that our data lends more support to the plus branch, corresponding generally to larger intra- than extra-axonal diffusivity. This was further backed by better numerical agreement to histologically measured (in rat spinal cord) axonal volume fractions for the plus branch. Finally, we compared the inferred time dependence of compartmental diffusivities to the predictions from effective medium theory (Burcaw et al., 2015; Novikov et al., 2014) with reasonable agreement.

*DKI metrics*

The observed time dependence of the net diffusivities was more pronounced than previous findings (Nilsson et al., 2009). One possible reason for this difference could be our usage of a very small diffusion gradient pulse width ($\delta = 1.15$ ms), in turn implying lesser blurring of spin trajectories (Mitra and Halperin, 1995). This may also play a role in the observation of nonconstant diffusivities all the way up to 350 ms, as well as the larger contrast in compartmental diffusivities across the ROIs. The observed nonmonotonic behavior of the kurtosis in some of the ROIs, also reported in (Pyatigorskaya et al., 2014), is a feature generally expected in nonuniform media. At very short times diffusion is Gaussian with a vanishing kurtosis. In the intermediate to long diffusion time regime with either nongaussian intracompartmental diffusion and/or heterogenous distribution of diffusivities across compartments, kurtosis is positive.

*Branch choice*

Although not fully resolved, we regard our time dependent diffusion data to provide more support to the plus branch than to the minus branch, corresponding here to intra-axonal diffusivity being larger than extra-axonal diffusivity in the spinal cord tissue. We base this on several observations:

First, the time-dependent behavior of intra-axonal diffusivity seems more plausible for the plus branch, decaying smoothly as function of diffusion time. The minus branch, on the other hand, was much noisier and even increased in some ROIs, which is unphysical. The plus branch compartmental diffusivities estimated by fitting to the biophysical model indicated intra-axonal diffusivity to be roughly a factor of 2 larger than the axial extra-axonal diffusivities in the fixed pig spinal cord white matter for the plus branch. A mostly larger intra-axonal diffusivity was also found in a recent in vivo study (Novikov et al., 2016b), and is also consistent with previous measurements in fixed rat brain utilizing very high b-values (Jespersen et al., 2010; Jespersen et al., 2007b). The value of 1.6 $\mu m^2$/ms found for water diffusion within the giant squid axon (Beaulieu and Allen, 1994) for diffusion time of 30 ms is also within the range of intra-axonal diffusivities found here. Except for one data point, all measurements fall below the diffusivity of water at the sample temperature.

Second, at early diffusion times, extra-axonal axial diffusivity tended to be much more similar to the extra-axonal radial diffusivity for the plus branch, in contrast to the minus branch where they differed substantially. As the diffusion time approaches zero, waters spins have had little chance to interact significantly with barriers, and diffusion is expected to appear increasingly isotropic.

Third, the estimated values for the axonal water fraction were in better agreement with histological measurements obtained from rat spinal cords in the same white matter tracts.

Finally, the finite axonal dispersion values found for the plus branch is in better correspondence to histologically obtained dispersion values for human spinal cord reported in (Grussu et al., 2016). Grussu et al make a number of estimates of dispersion in human spinal cord tissue based on different filter widths in their structure tensor analysis, as well as fitting to six different directional distributions, and it is therefore not possible to give a single number which characterizes axonal dispersion in white matter from their paper. Moreover, their analysis is based on 2D (sagittal) sections, which will likely tend to underestimate 3D dispersion. Nevertheless, we can make some rough estimates based on their Figure 5. They report median dispersions in white matter in terms of standard deviations of a 2D Gaussian distribution to be mostly between about 10 and 30 degrees. For their 2D Watson distribution, assuming that their reported circular variances result from a 2D projection of an underlying 3D Watson distribution, we estimate their median dispersions to correspond to 3D concentration parameters $\kappa$ between 2 and 21. For their filter width of one 1um, the corresponding 3D concentration parameters lie between 4 and 21. This is in good numerical agreement with our findings for the plus branch, and quite different from the minus branch, and a similar picture is seen in other highly coherent fibre bundles, such as the corpus callosum and the optical tract (Ronen et al., 2014; Schilling et al., 2016). Furthermore, $\kappa$ showed a decrease (increasing concentration parameter $\kappa$ with increasing diffusion time) as function of diffusion time, which is to be expected when the spins sample larger and larger segments of the axons. The dispersion for the minus branch on the other hand tended to be as large as possible, reflecting that the best solution to the equations is achieved for zero dispersion/infinite concentration parameter. This also means that strictly speaking, there is no exact solution for the minus branch, although the errors are quite small and much less than what one would typically achieve using least-squares optimization. Nevertheless, this behavior probably signifies that the one parameter Watson distribution does not precisely capture the full complexity of fODF in the spinal cord.

Evidence in favour of the plus branch has also been reported using isotropic diffusion encoding (Dhital et al., 2015; Lampinen et al., 2017). The observation of a small kurtosis from such sequences indicates a small difference between mean diffusivities in the intra-and extra axonal compartments: with a small to vanishing intra-axonal radial diffusivity, this is only possible if intra-axonal radial diffusivity is larger than extra axonal radial diffusivity. Conversely, in a recent study, the plus branch was found to yield intra-axonal diffusivity estimates comparable or even exceeding that of free water in human brain and rat brain in vivo, as well as rat spinal cord ex vivo (Hansen et al., 2017). The present analysis also yields unphysical $D_a$ estimates in the plus branch for one data point at the earliest diffusion time, and several values approaching those of the free diffusivity of water at 25°C. Adding the time-dimension is therefore crucial for resolving the branch selection problem as evident from the branch differences in $D_a$ behavior seen in Fig. 3, where physically sound time-behavior is only seen for the plus branch. Nevertheless, it is worth noting that branch selection may be tissue dependent (Novikov et al., 2016b), and could specifically differ between gray and white matter, ex vivo and in vivo, and across species.

On the other hand, we did find the plus branch radial diffusivity to be larger than axial diffusivity asymptotically (using EMT) in the extra-axonal space of about half of the ROIs. Although there is nothing fundamental preventing this to be the case, it is very surprising and also at variance with diffusion of TMA+ measured with iontophoresis (Prokopová et al., 1997). We suspect this finding to be a result of noise fluctuations at the largest diffusion times.

Nevertheless, weighing the evidence and in order to focus the subsequent analysis, we limited our comparison to effective medium theory to the plus branch. Likewise, we will concentrate the following discussion on this solution.

*Model parameters*

Extra-axonal axial diffusivity was generally less than half that of intra-axonal diffusivity, and showed a strong diffusion time dependence like its intra-axonal counterpart. Similar observations for the net axial diffusivity was made recently (Fieremans et al., 2016), and indicate that the hollow cylinder approximation of diffusion inside neurites is incomplete at clinically accessible diffusion times. The functional form of the time dependence of intra-axonal axial diffusivity was consistent with the predictions from effective medium theory, in the presence of Poissonian disorder, i.e. obstacles to diffusion with Poissonian spatial statistics. The biological nature of these obstacles is unknown at present, but candidates include axonal varicosities (Shepherd and Raastad, 2003; Shepherd et al., 2002), which is commensurate with the correlation lengths of ~1-10 μm found here (Table 1). Extra-axonal radial diffusivity approached the value of extra-axonal axial diffusivity for low diffusion times, when water molecules have not yet had time to interact significantly with the axonal membranes. The estimated intra-axonal water fraction was stable for the larger diffusion times, and had values in the range of 0.45 to 0.65 for the plus branch. This is in good agreement with the volume fractions estimated with histology from the same white matter tracts in rat spinal cord (Nunes et al., 2017), but substantially larger than the values of 0.2 to 0.4 from the minus branch. Although variations may exist between rat and pig, these differences are likely smaller than the differences between branch estimates. It must be noted however, that the MR estimated axonal volume fractions are affected by compartmental differences in transverse relaxation rates, which will tend to increase the effective intra-axonal volume fraction, which some have suggested has the larger $T_2$ (Bonilla and Snyder, 2007; Wachowicz and Snyder, 2002). The stability of the volume fractions at long diffusion times is consistent with the assumption of vanishing exchange of water across the fixated myelinated axons. For the shortest diffusion times, the axonal volume fraction estimate begins to depend systematically on diffusion time (Fig. 6). This is artifactual and probably due to model shortcomings, or bias in the estimation of the DKI parameters (Chuhutin et al., 2016), see below. However, there could potentially be a small effect of differential $T_1$ relaxation, as the diffusion time variation is achieved by varying the mixing time leading to variable longitudinal storage time.

*Limitations*

One assumption of the biophysical model is that intracompartmental kurtosis (specifically kurtosis in the intra-and extra-axonal compartments) can be ignored. Given our finding of a time-dependent intra-and extra-axonal diffusivity, implying nongaussian diffusion, this cannot be strictly true. Effective medium theory predicts the importance of intracompartmental kurtosis to decrease as coarse graining proceeds with increasing diffusion time, implying that the quality of the approximation should improve when the diffusion length become large compared to the correlation length (Novikov and Kiselev, 2010). Further studies are required in order to examine the influence of intracompartmental kurtosis, but given the correlation lengths and diffusivities reported in Table 1, we believe that it has a minor influence on the results when the diffusion time is larger than 20 ms. On the other hand, it may explain the overestimated intra-axonal diffusivity for short diffusion times.

Although our analysis crucially included axonal orientation dispersion, it was assumed to be axially symmetric and conform to the Watson distribution. This is likely to be a good approximation in spinal cord, although it may fail in regions with a substantial amount of collateral fibers, such as the cervical enlargement (Lundell et al., 2011). Most of our ROIs were placed away from the dorsal/ventral horns, where most collateral fibers emerge, and in a thoracic slice position. Furthermore, the ROIs almost fully fulfilled the threshold definitions in terms of diffusion tensor indices of linearity cL, sphericity cS, planarity cP, which have been suggested to identify voxels appropriate for white matter tract integrity analysis (WMTI), which has strong requirements to the alignment of fibers (Fieremans et al., 2011). Specifically, we found that at all diffusion times cL>0.4, cP<0.2, and cS<0.4. The threshold suggested by Fieremans et al. are cL $\geq 0.4$, cP $\leq 0.2$, and cS $\leq 0.35$. Thus only the sphericity threshold is not exactly fulfilled in all ROIs, but once the diffusion time exceeds 25 ms, cS<.35 in all ROIs. Likewise, the axially symmetric Watson distribution assumed here limits the application to specific regions in brain. Nevertheless, by adding dispersion it is reasonable to believe that it will increase the region of applicability as compared to the original WMTI, which can accommodate limited dispersion albeit without modelling it explicitly (Fieremans et al., 2011; Hansen et al., 2017).

We also neglected radial intra-axonal diffusivity (effectively equivalent to setting the axonal diameter to zero), which is expected to be a good approximation for most of the data with the diffusion acquisition parameters employed here and the dimensions of spinal cord axons (diameters <4 um (Neto Henriques et al., 2015)). However, for the shortest diffusion times and the largest b-values, there will be a non-negligible attenuation of intra-axonal spins ( $qa \sim 1$ ) for a diffusion gradient perpendicular to the largest axons in e.g. the reticulospinal spinal tract, and this may bias the corresponding estimates. We attempted to fit the full biophysical model in Eq. (6) with a radial intra-axonal diffusivity included, but this proved to be unstable. However, Monte Carlo simulations (data not shown) showed that the signal from spins within 2.83 µm diameter axons (table 1 average) perpendicular to the diffusion gradient attenuate less than the SNR for the largest b values once the diffusion time exceeds 35 ms.

Finally, only 2 types of compartments were included in the modelling, intra-axonal and extra-axonal water. As such, water residing between the myelin sheaths is ignored. Preliminary data on the same magnet indicates that this compartment has an effective T2 of 5-15 ms, meaning that its signal contribution will be subdominant compared to the other 2 compartments, but this may change at lower field strengths. Nevertheless, its omission here may affect the estimated parameters. Likewise, exchange between intra-/extra axonal water and myelin water was not taken into account, which may explain some of the time dependence observed here (Lee et al., 2017), although mainly at the longest diffusion times (Barta et al., 2015).

Our approach was based on estimating biophysical parameters from diffusion kurtosis parameters, which we found to give stable estimates with smooth diffusion time dependence, in contrast to direct nonlinear fitting to the signal. However, errors in the estimation of diffusion kurtosis parameters due to both noise and the influence of higher order cumulants are inevitable (Chuhutin et al., 2016), and will propagate into the estimated model parameters. To mitigate this, we acquired six b shells, and preprocessed the data using denoising and correction for Gibbs ringing.

Finally, we note that one must be cautious generalizing our results to the in vivo case, as tissue death, fixation and temperature could have profound influences on tissue microstructure.

*Comparison to EMT*

We found varying agreement to the functional form of the time dependence of the diffusivities predicted by Novikov et al. (Burcaw et al., 2015; Fieremans et al., 2016; Novikov et al., 2014) using effective medium theory, showing that the spatial correlations in the disorder (obstacles for diffusion) determine the approach of the diffusivities to the long time asymptotes. Intuitively this is understood by dividing the tissue into slabs with dimensions on the order of the diffusion length. At some point when the diffusion time is large enough, the slabs are statistically similar, and diffusivity will no longer depend on time. The rate at which these slabs become representative of the disorder depends on the nature of the fluctuations, quantified by the 2-point correlation function describing the spatial range of fluctuations, which therefore defines the approach to the tortuosity limit (Novikov et al., 2014). Our results, especially for intra-axonal diffusivity, indicate that this disorder is Poissonian, which was also found in prior studies (Burcaw et al., 2015; Fieremans et al., 2016;

Novikov et al., 2012). Taking this time dependence into account was recently shown to be important for accurate estimation of e.g. axonal densities in human brain (De Santis et al., 2016). The extra-axonal diffusivities showed more variable agreement to EMT, although no systematic deviations could be discerned. Further investigations of the agreement to EMT should therefore be a focus of future studies.

## Conclusion

Covering an extensive range of diffusion times on a 16.4T microimaging system, we demonstrated strong time dependence of diffusion and kurtosis in well-defined anatomical white matter tracts in fixed spinal cord. Using biophysical modeling, this was related to microstructural parameters including axonal water fraction, dispersion, and intra-and extra-axonal diffusivities. We found evidence that in fixed porcine spinal cord white matter, intra-axonal diffusivities are substantially larger than the extra-axonal diffusivity, both depending substantially on diffusion time. A reasonable correspondence to effective medium theory for the time-dependent diffusivities was demonstrated. Our results are important for further progress of microstructural modelling, as well as proper interpretation of diffusion measurements.

## Acknowledgements

The authors were supported by the Danish Ministry of Science, Technology and Innovation's University Investment Grant (MINDLab, Grant no. 0601-01354B). The authors acknowledge support from NIH 1R01EB012874-01 (BH), Lundbeck Foundation R83-A7548 (SNJ). The 9.4T lab was funded by the Danish Research Council's Infrastructure program, the Velux Foundations, and the Department of Clinical Medicine, AU. The 3T Magnetom Tim Trio was funded by a grant from the Danish Agency for Science, Technology and Innovation. NS would like to acknowledge support from the European Research Council (ERC) under the European Union's Horizon 2020 research and innovation programme (grant agreement No. 679058 - DIRECT-fMRI), as well as under the Marie Sklodowska-Curie grant agreement No 657366. The authors thank Dr. Daniel Nunes for assistance and discussions.

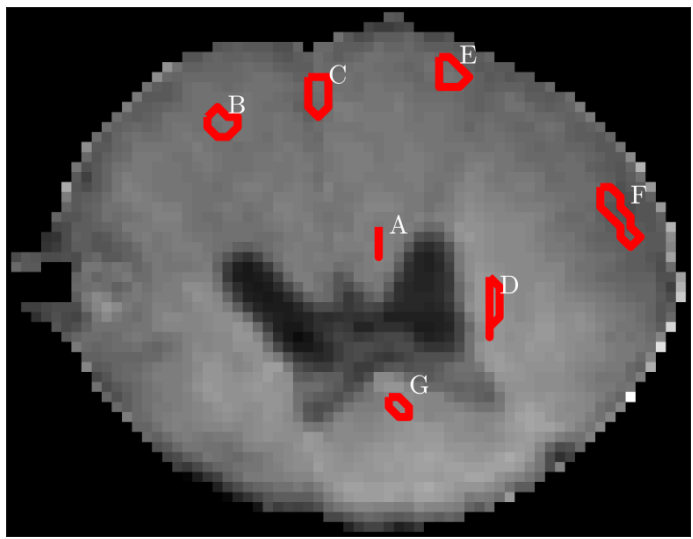

**Figure 1:** Definition of analyzed ROIs shown on a b=0 image. The outlined tracts are: A=dorsal corticospinal tract, B= Fasiculus Gracilis, C=Fasiculus Cuneatis, D=Reticulospinal tract, E=Rubrospinal tract, F=Spinothalamic tract, G=Vestibulospinal tract. These ROIs have been chosen according to the anatomical regions investigated in (Ong and Wehrli, 2010; Ong et al., 2008), reflecting a range of axonal radii and densities.

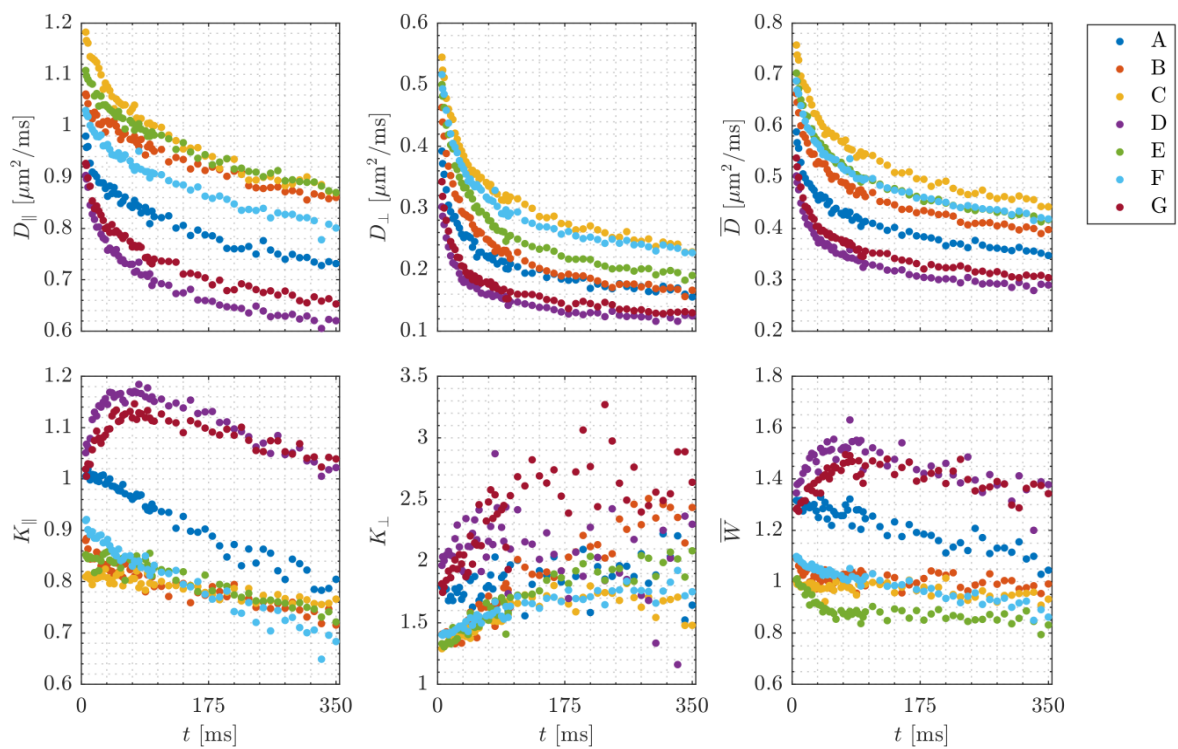

**Figure 2:** Diffusion time (t) dependence of net diffusion tensor (top row) and kurtosis metrics (bottom row) in the 7 ROIs as determined by nonlinear fitting to axisymmetric DKI. Note that mean

diffusivity is a linear combination of axial and radial diffusivities, in contrast to mean kurtosis, which is independent of axial and radial kurtosis.

$$S \Big/ S_{rep}$$

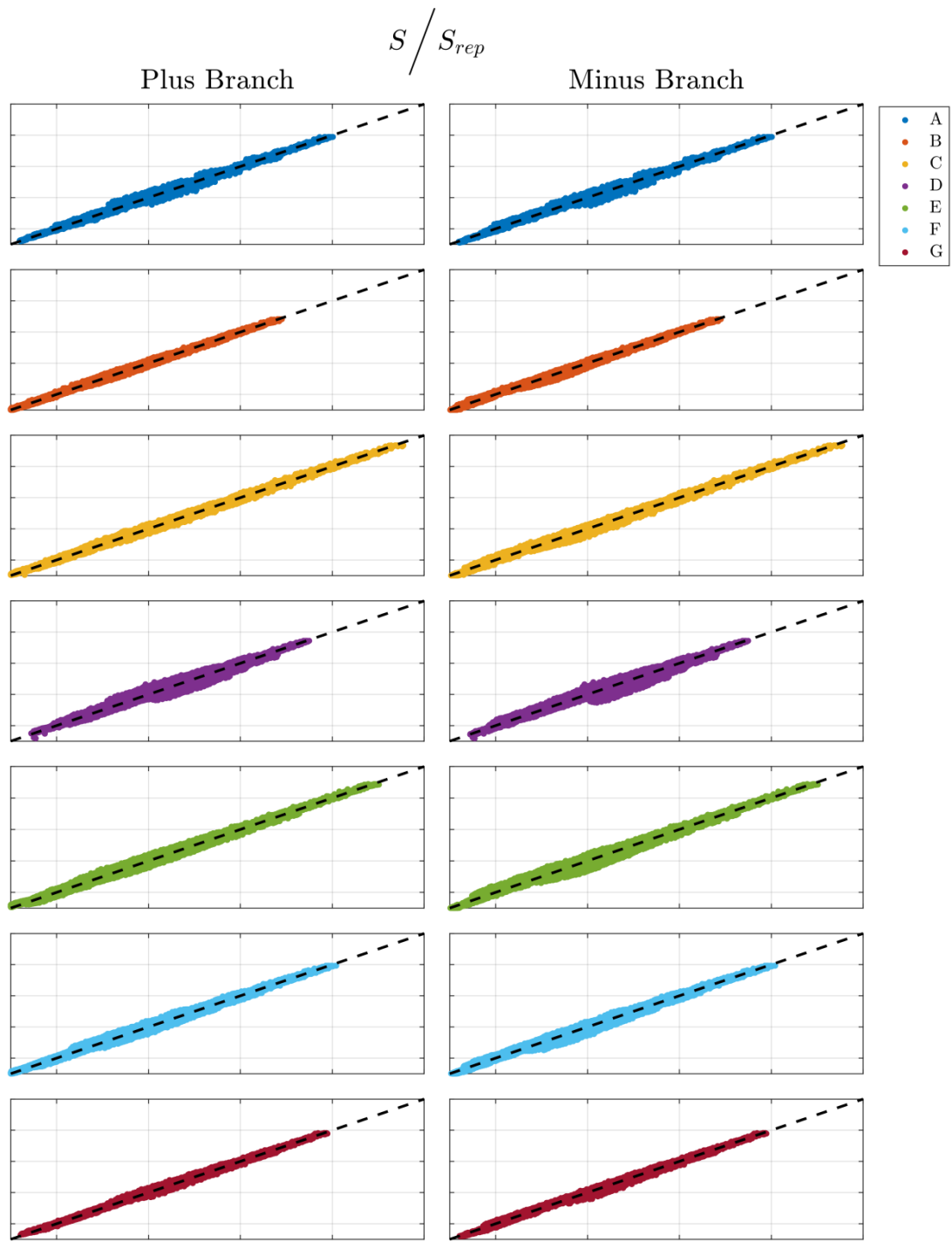

**Figure 3:** Acquired signal (y-axis) versus predicted signal (x-axis) using the estimated model parameters and the biophysical model. Each of the graphs contain data from ROI, and each of the points correspond to one voxel from one diffusion-weighted image, i.e. a given value of diffusion weighting, diffusion direction, diffusion time.

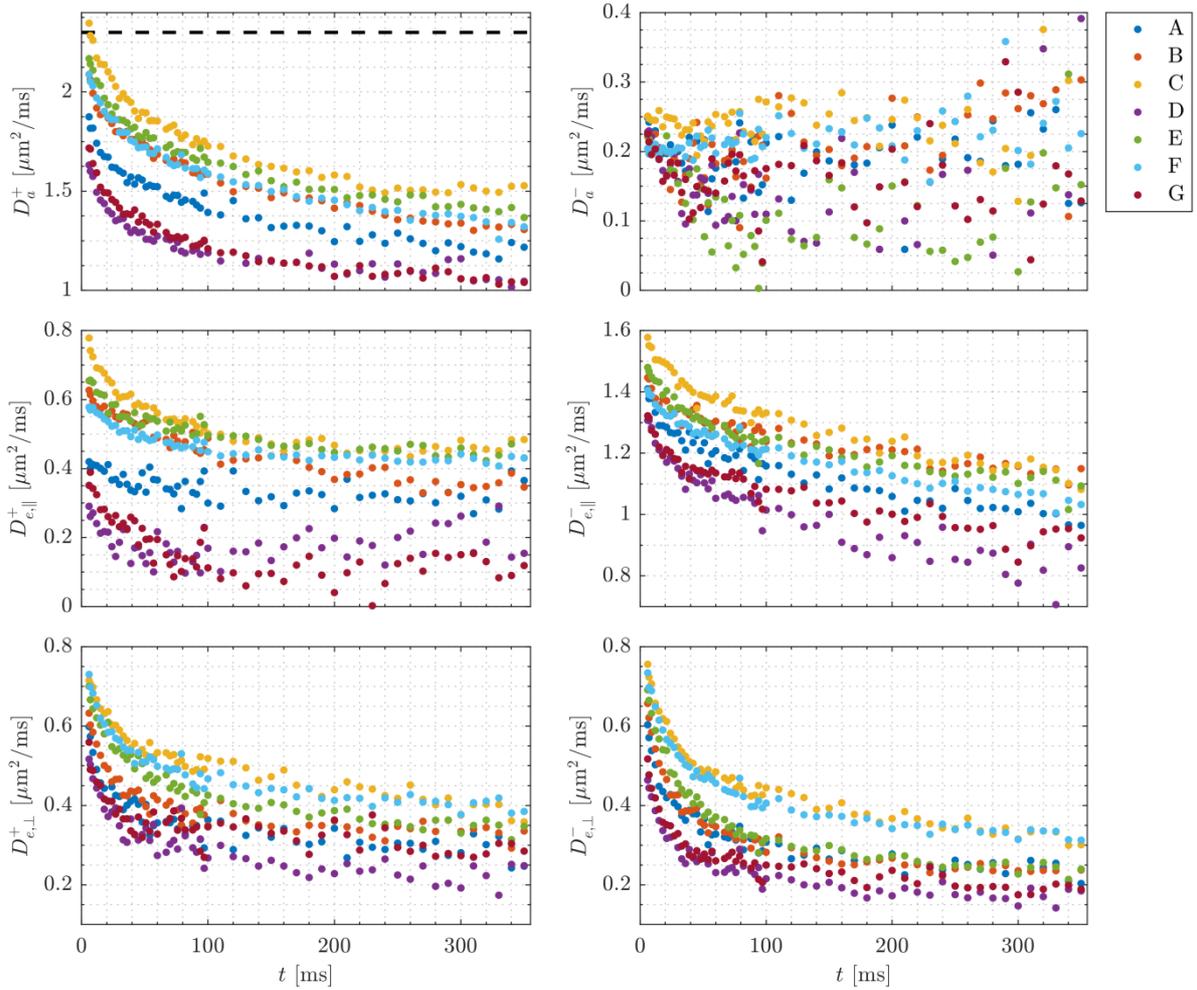

**Figure 4:** Intra-axonal diffusivity (top row) as function of diffusion time for the 2 branches, denoted by superscripts + or - according to which sign was chosen in the solution to the underlying quadratic equation. The dashed horizontal line marks the diffusivity of free water at the sample temperature. Middle row depicts extra-axonal axial diffusivity as function of diffusion time for both solutions, and bottom row extra-axonal radial diffusivity as function of diffusion time for both solutions.

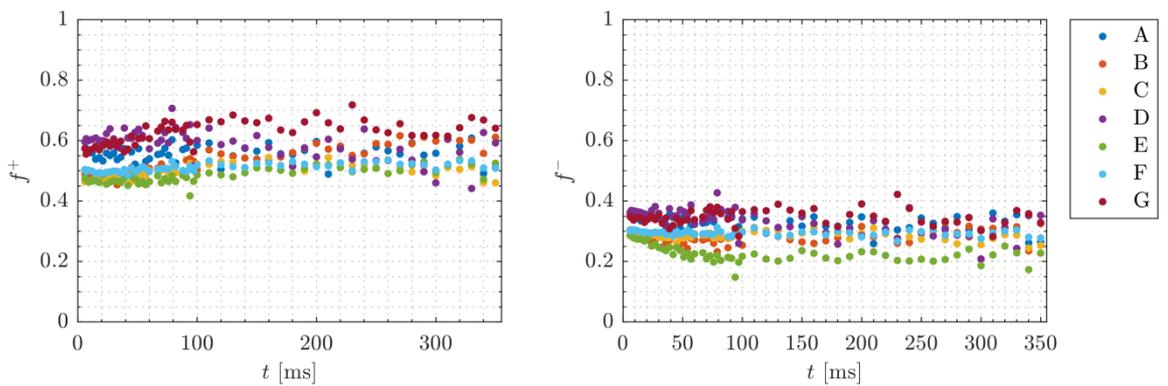

**Figure 5:** Axonal volume fraction for both solutions. Assuming no exchange across the axonal membrane, and no $T_1$ effects, they should be constant as a function of diffusion time.

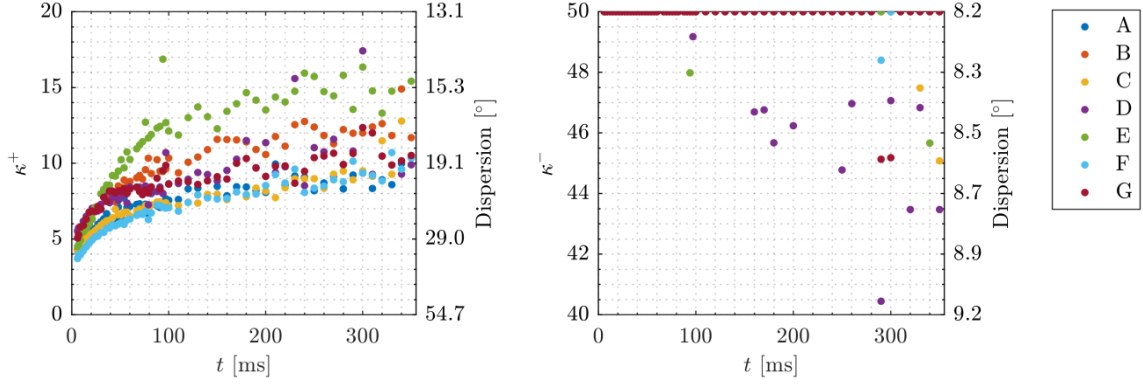

**Figure 6:** Concentration parameter $\kappa$ of the Watson distribution. Also shown on the right y-axis is the corresponding dispersion measured in degrees, defined by $\arccos\left(\sqrt{\langle(\hat{c}\cdot\hat{u})^2\rangle}\right)$, i.e. essentially the angular variance.

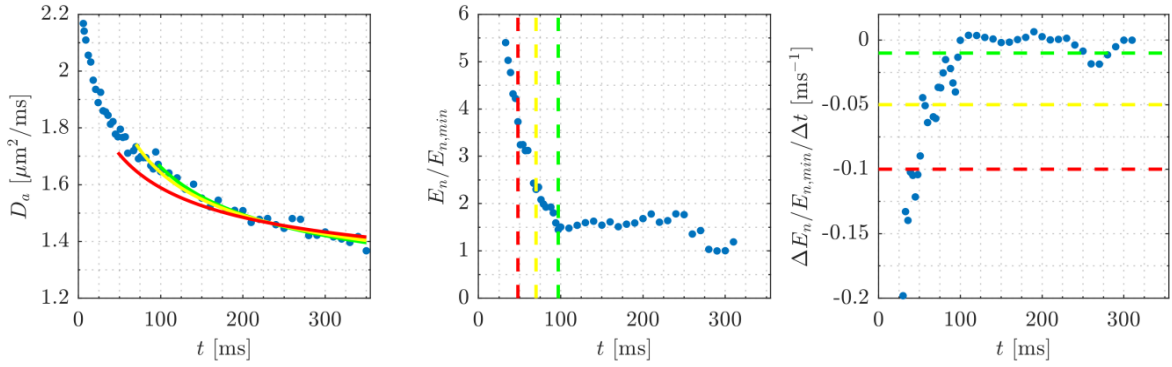

**Figure 7:** Example of the fitting procedure used for fitting to effective medium theory using intra-axonal diffusivity from the spinothalamic tract. Fitting has been performed from a variable lower diffusion time, the cutoff time, up to the largest diffusion time (350ms). The left panel shows intra-axonal diffusivity from the spinothalamic tract (blue circles) along with 3 example fits for varying ranges of diffusion time t (solid lines). The middle panel shows the mean square error $E_n$, normalized by the minimum mean square error $E_{n,\min}$, as function of the cutoff time, and the dashed vertical lines mark the cutoff times used for the respective fits. The right panel shows the selection of these 3 cutoff times. The y-axis corresponds to the rate of change of the normalized mean square error, i.e. the slope of the data in the middle panel: when the fitting range extends to too low diffusion times, the magnitude of mean square error increases sharply. The green, yellow and red lines correspond to the point at which the derivatives passed the 0.01, 0.05, and 0.1 limits, as marked by the dashed lines.

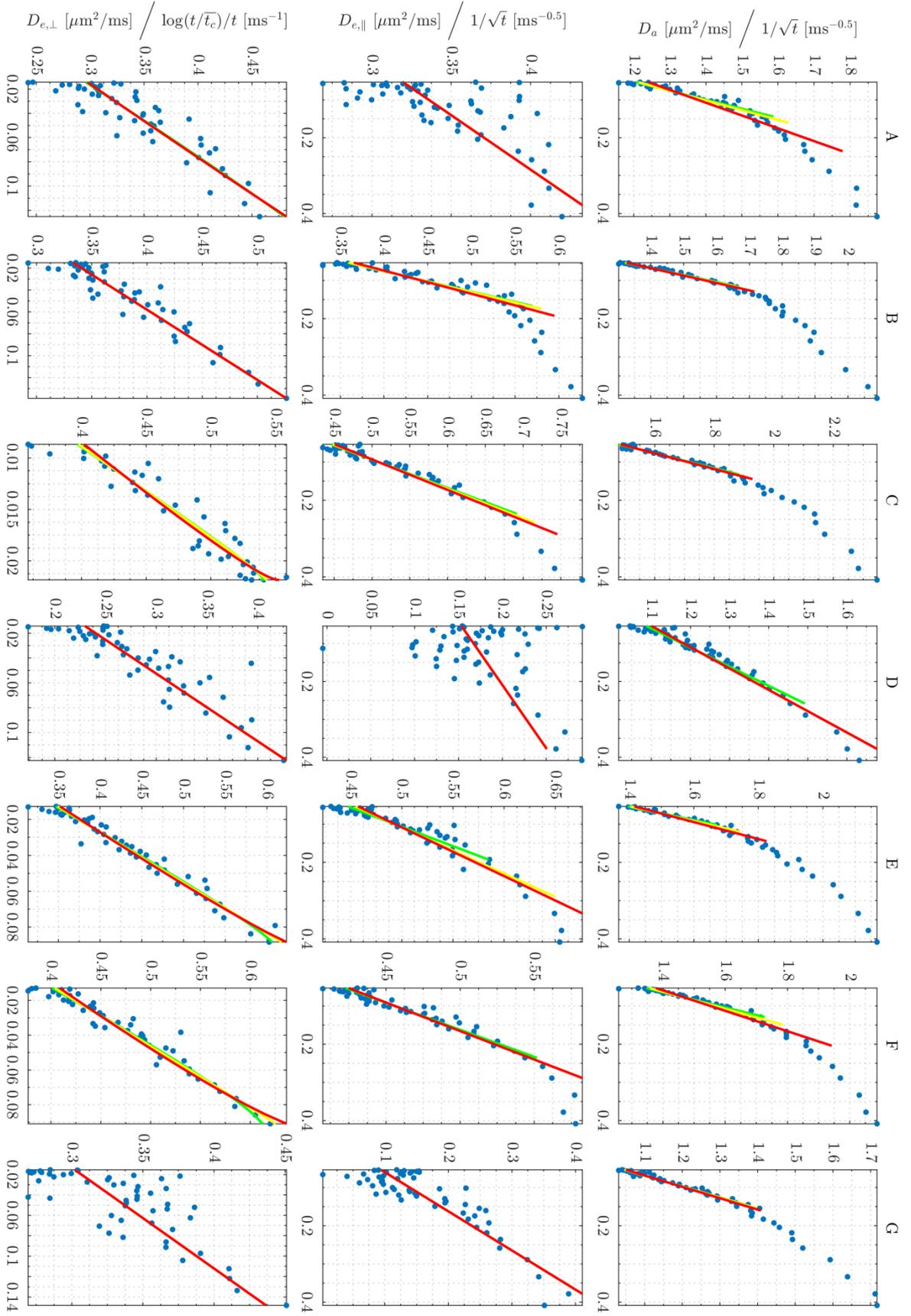

**Figure 8:** Comparison to effective medium theory for the plus branch for all ROIs and all diffusivities. Each of the diffusivities correspond to one row as listed on the left-hand side, and each column corresponds to one ROI as indicated on the top.

**Table 1:** Fitting parameters for the diffusivities from effective medium theory Eq. (16) in each of the investigated white matter ROIs. Diffusivities are in μm²/ms and $t_c$ in ms , whereas the units of the fitting constants $c_{1\text{-}3}$ follow from Eq. (16). Also shown are cutoff times in ms defining the lower limit of the fitting range. Correlation lengths $l_c$ in μm have been determined by $\sqrt{4D_{e,\perp}t_c}$ for the extracellular radial case, $c_2\sqrt{\pi/D_{e,\parallel}(\infty)}$ for the extracellular axial case, and as $c_1\sqrt{\pi/D_a(\infty)}$ for the intra-axonal $l_c${Fieremans, 2016 #1821}. The final 2 rows are from rat spinal cord and were obtained from (Nunes et al., 2017).

| | | A | B | C | D | E | F | G |
|---|---|---|---|---|---|---|---|---|
| $D_a$ | $D_a(\infty)$ | 0.98 | 1.03 | 1.21 | 0.98 | 1.09 | 1.08 | 0.86 |
| | $c_1$ | 4.19 | 5.52 | 5.13 | 1.98 | 5.73 | 5.01 | 3.53 |
| | cutoff | 16.00 | 24.00 | 19.00 | 5.00 | 30.00 | 20.00 | 16.00 |
| | $l_c$ | 7.50 | 9.67 | 8.28 | 3.54 | 9.73 | 8.55 | 6.76 |
| $D_{e,\parallel}$ | $D_{e,\parallel}(\infty)$ | 0.30 | 0.26 | 0.37 | 0.14 | 0.39 | 0.39 | 0.04 |
| | $c_2$ | 0.35 | 1.87 | 1.35 | 0.30 | 0.99 | 0.71 | 0.98 |
| | cutoff | 2.00 | 12.00 | 6.00 | 2.00 | 9.00 | 6.00 | 2.00 |
| | $l_c$ | 1.12 | 6.50 | 3.91 | 1.43 | 2.79 | 2.02 | 8.48 |
| $D_{e,\perp}$ | $D_{e,\perp}(\infty)$ | 0.27 | 0.31 | 0.29 | 0.20 | 0.30 | 0.36 | 0.29 |
| | $c_3$ | 1.76 | 1.56 | 12.32 | 1.84 | 3.93 | 3.32 | 1.01 |
| | $t_c$ | 2.53 | 1.87 | 18.31 | 2.39 | 4.62 | 4.81 | 1.62 |
| | cutoff | 3.00 | 2.00 | 14.00 | 2.00 | 4.00 | 5.00 | 2.00 |
| | $l_c$ | 1.66 | 1.52 | 4.63 | 1.40 | 2.35 | 2.63 | 1.36 |
| Tortuosity | $D_{e,\parallel}(\infty)/D_{e,\perp}(\infty)$ | 1.11 | 0.84 | 1.28 | 0.67 | 1.32 | 1.07 | 0.15 |
| Diameter (rat) | μm | 1.43 | 2.29 | 3.39 | 4.09 | 2.66 | 2.53 | 3.43 |
| Axon Vol. Frac. (rat) | | 0.57 | 0.55 | 0.47 | 0.38 | 0.47 | 0.42 | 0.52 |